# Reduction of breast cancer relapses with perioperative non-steroidal anti-inflammatory drugs: new findings and a review

Running title: Reduction of breast cancer relapses with perioperative NSAID

Michael Retsky[1,2,*]
Email: michael.retsky@gmail.com                              [*] Corresponding author.

Romano Demicheli[3]
Email: romano.demicheli@istitutotumori.mi.it

William JM Hrushesky[4]
Email: william.hrushesky@gmail.com

Patrice Forget[5]
Email: forgetpatrice@yahoo.fr

Marc De Kock[5]
Email: marc.dekock@uclouvain.be

Isaac Gukas[6]
Email: igukas@hotmail.com

Rick A Rogers[1]
Email: rogers@bioimage.harvard.edu

Michael Baum[7]
Email: michael@mbaum.freeserve.co.uk

Vikas Sukhatme[8]  Email: vsukhatm@bidmc.harvard.edu

Jayant S Vaidya[9]
Email: jayantvaidya@gmail.com

[1] Harvard School of Public Health, BLDG I, Rm 1311, 665 Huntington, Ave, Boston, MA 02115, USA

[2] Royal Free and UCL Medical School, Centre for Clinical Science and Technology, University College London, Clerkenwell Building, Archway Campus, Highgate Hill, London, UK

[3] Scientific Directorate, Fondazione IRCCS Istituto Nazionale Tumori, Via Venezian 1, 20133 Milan, Italy

[4] Oncology Analytics, Inc, 8751 W. Broward Blvd, Suite 500, Plantation FL 33324, USA

[5] Department of Anesthesiology, Universite Catholique de Louvain, St-Luc Hospital, av. Hippocrate 10-1821, 1200 Brussels, Belgium

[6] James Paget University Hospital, Lowestoft Road, Gorleston, Great Yarmouth, Norfolk NR31 6LA, UK

[7] Royal Free and UCL Medical School, Centre for Clinical Science and Technology, University College London, Clerkenwell Building, Archway Campus, Highgate Hill, London N19 5LW, UK

[8] Beth Israel Deaconess Medical Center, 330 Brookline Ave, Boston, MA 02215, Harvard Medical School, and GlobalCures, Boston, MA 02115

[9] Clinical Trials Group of the Division of Surgery and Interventional Science, University College London, Clerkenwell Building, Archway Campus, Highgate Hill, London N19 5LW, UK

Abstract

To explain a bimodal pattern of hazard of relapse among early stage breast cancer patients identified in multiple databases, we proposed that late relapses result from steady stochastic progressions from single dormant malignant cells to avascular micrometastases and then on to growing deposits. However in order to explain early relapses, we had to postulate that something happens at about the time of surgery to provoke sudden exits from dormant phases to active growth and then to detection. Most relapses in breast cancer are in the early category. Recent data from Forget et al suggests an unexpected mechanism. They retrospectively studied results from 327 consecutive breast cancer patients comparing various perioperative analgesics and anesthetics in one Belgian hospital and one surgeon. Patients were treated with mastectomy and conventional adjuvant therapy. Relapse hazard updated Sept 2011 are presented. A common Non-Steroidal Anti-Inflammatory Drug (NSAID) analgesic used in surgery produced far superior disease-free survival in the first 5 years after surgery. The expected prominent early relapse events in months 9-18 are reduced 5-fold.  If this observation holds up to further scrutiny, it could mean that the simple use of this safe, inexpensive and effective anti-inflammatory agent at surgery might eliminate early relapses. Transient systemic inflammation accompanying surgery could facilitate angiogenesis of dormant micrometastases, proliferation of dormant single cells, and seeding of circulating cancer stem cells (perhaps in part released from bone marrow) resulting in early relapse and could have been effectively blocked by the perioperative anti-inflammatory agent.

Key words

Breast cancer, Early relapse, Late relapse, Angiogenesis, Inflammation, Surgery, Analgesic, Perioperative, Metastases, Bleeding, NSAID, dormancy

*Introduction*
**Historical overview**
The history and philosophy of science describes progress not simply in steady incremental steps but with rare and welcome sudden leaps forward. Karl Popper described the hypothetico-deductive process of observation and experimentation as "normal science" [1] whereas Thomas Kuhn described the occasional leap forward as "revolutionary science" and coined the expression "paradigm shift" to describe this phenomenon [2]. Normal science demands a method but revolutionary science demands an open mind. The recent history of the search for the cure for breast cancer can be described in this way.

From about 200 AD until the early 19thC, breast cancer was treated according to the traditions of the Galenic doctrine that declared breast cancer as a systemic disorder resulting from an imbalance of the natural "humours" with excess of the putative black bile. Relative to his time Galen (131-203AD) was a true scientist however this was a purely metaphysical construct that led to barbaric therapeutic interventions aimed at restoring the balance of the "humours" [3]. Vesalius (1514–1564) was one of the first to vigorously oppose Galen's doctrines. Yet, the first scientific revolution in the history of breast cancer was the description by Virchow (1821-1902) of the cellular nature of cancer and its propensity to spread along the lymphatic system to be arrested in the axillary lymph nodes [4]. This mechanistic concept informed the evolution of the radical mastectomy.

William Halsted (1852-1922) operated at a time when the triumph of mechanistic principles was at its peak when the common man had begun enjoying the fruits of the Industrial Revolution. His surgical expertise was remarkable, and for the first time, breast cancer seemed curable with local recurrence rates of only 10% at 3 years, very low compared to the other series at that time. Unfortunately, only about a quarter of patients treated by Halsted survived 10 years [5]. Thus even when the tumor seemed to have been completely 'removed with its roots', the patients still developed distant metastases and succumbed with little evidence of "cure" if patients were followed up for as long as 25 years [6]. From the popularization of the classical radical mastectomy at the very end of the 19thC until about 1975, almost all patients with breast cancer, of a technically operable stage, were treated with modifications of the radical mastectomy.

**The biological revolution of the late 20thC.**
Prompted by the failures of radical operations to cure patients with breast cancer, Bernard Fisher proposed a revolutionary hypothesis that rejected the mechanistic model of the past replacing it with a biological model which challenged and refuted every prediction from the time of Virchow [7]. H postulated that cancer spreads via the blood stream bypassing the lymphatic channels and that this can occur even before the lump is first detectable; the rate of growth and the rate of spread being determined by the nature of the malignant focus at its inception.

There were two therapeutic consequences of this conceptual revolution:
(A) The extent of local treatment by surgery and radiotherapy might control the disease on the chest wall but have no effect on survival, the horse (cancer) having bolted before the stable door (radical surgery) was slammed shut.
 (B) If the outcome of treatment was pre-determined by the extent of microscopic subclinical metastases present at the time of diagnosis, then the only chance of cure would be with adjuvant systemic therapy i.e., drugs targeting these putative sites of disease, even for patients apparently only with localized tumors.

As regards the extent of local treatment, many trials have tested less versus more surgery with or without post-operative radiotherapy [7,8]. The extent of surgery doesn't appear to have significant impact on survival. The second prediction following the conceptual revolution spearheaded by Fisher, has been spectacularly corroborated by the success of the trials of adjuvant systemic therapy that have been mirrored by the fall in breast cancer mortality in many parts of the world since the mid-1980s [9,10].

Progress has now slowed down and is now best described in small incremental steps that begin to suggest an exhaustion of the contemporary paradigm [11, 12]. Unfortunately revolutionary thinking cannot be conjured on demand; the *zeitgeist* has to be right. We believe that the time is indeed right for the rebuilding of a conceptual model of breast cancer resulting from the identification of outlying observations that are incapable of explanation according to the Fisherian thinking in its present form. It is our task to develop a bold set of conjectures that together have greater explanatory power while at the same time accounting for the undoubted successes of the last 40 years.

**Toward a new understanding of the natural history of the disease.**
Among the most striking inconsistencies between the "Fisherian" model and clinical observations, is the pattern of hazard rates for local and distant recurrences after surgery for clinically localized disease. Instead of these demonstrating a shape that would be consistent with a stochastic pattern of transition from sub-clinical micro-metastases at different stages of progression and different rates of cellular proliferation, we witness a double peak, the first a steep and narrow based peak at about one or two years after surgery and a second lower and wider based curve reaching its plateau at about five or six years. [See figures 1 and 2.] These observations, repeated in almost every data set examined by smoothed hazard rate plots, cannot be explained by a linear dynamic implicit in the current conceptual model of breast cancer [13-17].

If the facts don't fit the model then the model is wrong, not the observations. Or in the words of Nassim Nicholas Taleb, "The black swan is an outlier, as it lays outside the realm of regular expectations….it carries an extreme impact and in spite of its outlier status, human nature makes us concoct explanations for its occurrence after the fact, making it explainable and predictable" [18].

Using non-linear (chaos theory) models, an adequate explanation can be found for the "black swans" that include the biphasic relapse pattern but at the same time can account for the undoubted successes of the contemporary paradigm. Although the number of metastases that are seeded by the primary tumor would be, at least as a working hypothesis, linearly related to the tumor size and biological aggressiveness, we suggest that the clinical appearance of metastases is often triggered or accelerated only after the primary tumor has been perturbed or removed. One has to assume that the majority of metastases at the time of diagnosis are dormant rather than actively growing. Within the "dormant" metastases, we can conceive of single quiescent isolated tumor cells and, moreover, others where there is some type of balance between cell growth and cell death. The latter may be partly determined by factors that inhibit angiogenesis without which a clump of cancer cells cannot grow to more than $10^6$ or $10^7$ cells in number and other factors that inhibit epithelial proliferation or encourage apoptosis. Immune related factors may also be involved. If stimulating factors are increased or inhibiting factors are reduced, the dormant condition can no longer be maintained [19-21].

It is well documented in animal models and humans that removal of the primary tumor can reduce the inhibition of angiogenesis and it is recognized that following surgery, there is a surge in cytokine production that promotes angiogenesis and growth factors aiding wound healing [15,22,23]. Thus it is not surprising that tumor angiogenesis and proliferation may be provoked by the surgery involved in the attempt to control primary cancer. Thus a likely trigger for 'kick-starting' the growth of dormant metastases could be the act of surgery itself. In support of this thesis is the observation that a wound response gene expression signature can predict breast cancer survival. [24]

After surgery for breast cancer, the first peak in the incidence of secondary disease occurs at about 1-2 years irrespective of whether the tumor was at stage I or stage III [25]. It is only the height of the peak that changes with stage, the later the stage at presentation the higher is the peak, ***but the timing of the signal remains the same***. These phenomena suggest a nonlinear dynamic model for breast cancer, which, like all chaotic systems, is determined by initial conditions around the time of diagnosis [26].

**Therapeutic consequences**
The therapeutic consequences of the new model are self-evident. Assuming that the primary surgery (mastectomy or lumpectomy) removes all macroscopic evidence of disease, we can then visualize two families of subclinical residual foci, either in the tumor bed, or nesting in distant organs. One group would comprise organelle like structures existing in a relatively unstable state of dynamic equilibrium, perhaps awaiting a kick-start following the surgical onslaught to a phase of active progression; and the other consisting of small clusters of dormant cells destined to become apparent at any time between one and 25 years after surgery and relatively unresponsive to the initial systemic response to the surgical trauma. It is possible that one therapeutic option if timed correctly might favorably impact on both families of subclinical foci but perhaps we need to consider different therapeutic interventions for each of the postulated residual foci of disease.

This then might comprise a complex schedule of a pre or peri-operative biological inhibition of the response to surgery over a short period of time followed by a long-term drug schedule over many years that could include, for example, endocrine agents for the hormone receptor positive group of diseases [27] or "personalized" systemic therapies for the hormone receptor negative group [28]. The peri-operative therapy, which is non-specific for the tumor's biological characteristics, could impact prominently on tumors displaying high rates of early relapse [29].

**Bimodal relapse pattern details.**
As noted, our analysis of data from the Milan National Cancer Institute found an unexpected bimodal pattern of relapse hazard among 1173 early stage breast cancer patients treated by mastectomy. Figure 1 shows Milan data for premenopausal patients and fig. 2 shows postmenopausal patients in relapse hazard format.

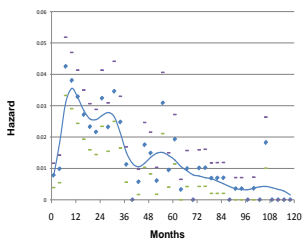

**Fig 1** Hazard of relapse for premenopausal patients treated at *Istituto Nazionale Tumori* in Milan, Italy. Hazard is the number of events that occur in a time interval divided by the number of patients who enter that time as event free. Patients were treated by mastectomy well before the routine use of adjuvant therapy. The time interval in all hazard figures used here is 3 months. Average and standard deviations are indicated as diamonds and bars. The curve was obtained by a kernel-like smoothing procedure.

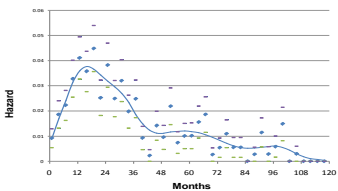

**Fig 2** Same as fig. 1 except that these are postmenopausal patients.

There is an early peak of recurrence risk during the first three years of follow-up, a nadir at 50 months and a broad second peak extending from 60 months to over 15 years. Fifty to eighty percent of relapses, the proportion increasing with primary tumor size, reside within the first peak. Under closer examination, the first peak consists of two distinct groups centered at 10 months and 30 months that are well distinguishable in premenopausal patients but occur for postmenopausal patients as well. This pattern was not explainable by accepted theories.

Similar patterns have now been identified in 20 independent databases from US, Europe and Asia. One of these databases is shown in Fig 3 in disease free survival format. It may be compared to Milan data shown in the same format in Fig. 4. This effect is apparently not restricted to breast cancer as we have noted case reports or similar recurrence dynamics among patients who are resected for primary control of prostate, lung, and pancreatic cancers, as well as osteosarcoma and melanoma [30-36].

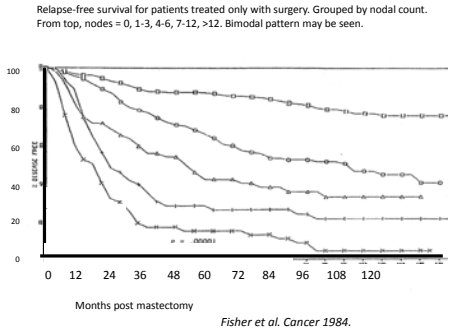

Fisher et al. Cancer 1984.

Fig 3. These are independent data for mastectomy treated patients in disease free survival format as modified from Fisher et al Cancer 1984 [37]. Patients are grouped by axillary lymph nodes invasion, with axillary lymph node tumor-free (N-) patients in the uppermost curve and patients with 10 or more invaded lymph nodes (N+) in the lower curve. The magnitude of the early relapse component may be visualized. For N- patients, after surgery alone 80% of patients are long lasting disease-free with half the relapses early and half late. For the poor prognosis patients (N+ with 10 or more involved lymph nodes), surgery alone is quite ineffective. Most of relapses (90%) are early and support the label of "poor prognosis". These data are visually and quantitatively very similar to the Milan data shown in fig. 4. Poor prognosis appears to have less impact on late relapses.

.

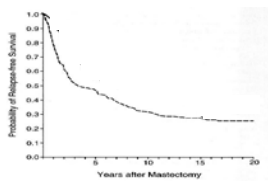

Bonadonna, Valagussa et al. NEJM 1995

Fig 4. Milan data for patients treated with mastectomy are presented in the more conventional format as disease free survival. The percentage disease free starts at 100% and rapidly drops until approximately 4 years where there seems to be a short plateau. Relapses start to happen again at about 5 years and slowly continue thereafter tapering off gradually at about 15 years. The plateau at 4 years corresponds to the end of the early relapses seen in figs. 1 and 2. Modified from Bonadonna et al NEJM 1995 [38].

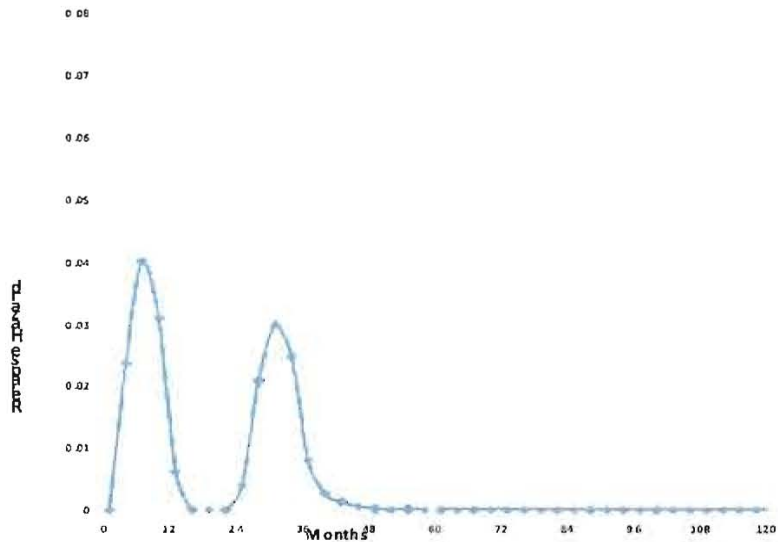

**Fig. 5** Computer simulations of early relapse events. Hazard of relapse for early events centered at 10 months and at 30 months post-surgery as proposed by computer simulation are shown. Simulations included effects of mastectomy and were based on Milan data shown in figs. 1 and 2. The 10 month and 30 month events may be distinguished in fig. 1 and are less clear but present in fig. 2.

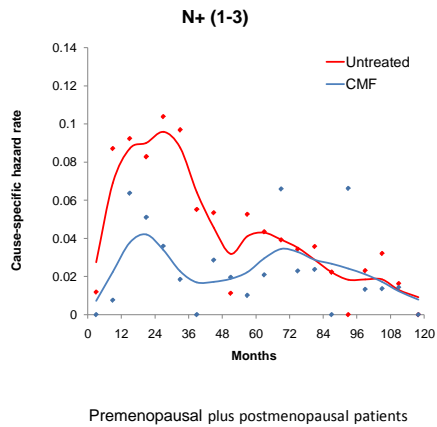

Premenopausal plus postmenopausal patients

**Fig. 6** The result of adjuvant CMF chemotherapy. The two early peaks in the untreated population coalesce into a single peak in this treated population at about 20 months. Apparently CMF chemotherapy acts to produce most extensive reduction in relapse hazard in the 1$^{st}$ and 3$^{rd}$ year [39].

The proposed model was further supported by the fact that in patients given adjuvant chemotherapy targeting proliferating cells first year (angiogenesis switching-related) recurrences and third year (single cell wake up-related) recurrences were remarkably reduced. Indeed, additional data from Milan for patients treated with CMF adjuvant chemotherapy (fig. 6) provided evidence that the recurrence risk pattern of patients receiving adjuvant chemotherapy displays a single initial peak at 18-20 months and a late peak at about 60 months.

**Most important finding – early relapses are the result of something that happens at surgery.**
The most important finding of this early work is that something happens at or about the time of surgery to accelerate or induce metastatic activity that results in early relapses. These early relapses comprise over half of all relapses. Surgery-induced angiogenesis of dormant avascular micrometastases and surgery-induced activity of single malignant cells are implicated. Late relapses are apparently not accelerated by surgery but the shallow peak at 5 years occurs as a result of shedding from primary ceasing after primary removal. We have been vigilantly looking for new data with which we can learn more about surgery-induced tumor activity and that perhaps will also lead to improved outcomes. As we describe here, there has been an important development.

*Materials and Methods*

**Forget et al data.**
In June 2010, Forget et al reported data from a retrospective disease free survival study of 327 consecutive patients treated in one Belgian hospital. Patients were compared according to the perioperative analgesics administered (sufentanil, clonidine, ketorolac and ketamine), following the preferences of the two anesthesiologists in charge [40]. As stated in the initial report, the sample size was limited by availability of medical records and to maintain oncologic treatment homogeneity.

Approval of the Ethical Committee of St-Luc Hospital was provided by the CEBH of the Université Catholique de Louvain (Brussels, Belgium), Chairperson Prof. Dr. J.M. Maloteaux. Investigators were unable to obtain consent from the patients for this retrospective study and the need for written informed consent from participants was waived, as accepted by the CEBH.

Patients with previous ipsilateral surgery for breast cancer were excluded. Indications for mastectomy with axillary clearance were defined according to international recommendations and guidelines [41,42]. These indications were discussed every week by the multidisciplinary board of the breast clinic and regularly updated and adjusted with new international recommendations and relevant literature. All mastectomies were performed by one surgeon and jointly followed by the surgeon and one oncologist. Chemotherapy, radiotherapy and endocrine therapy were performed according to the international expert consensus ($9^{th}$ and $10^{th}$ St-Gallen consensus) [43-46]. During the first two postoperative years, medical consultation occurred each three months, then every 6 months during three additional years and once a year thereafter.

Follow-up in that initial report was average 27.3 months with range 13-44 months. Patients who received anti-inflammatory drugs were compared with those who had not and their hazard of recurrence was analyzed and compared.

The type and the dosages of intraoperative analgesics used were, for sufentanil from 0 to 0.5 µg.kg-1, for clonidine from 0 to 6 µg.kg-1 (preincisional), for ketamine from 0 to 0.5 mg.kg-1 (preincisional). Ketorolac, when administered, was used as follow: 20 mg preincisional in patients under 60 kg, and 30 mg in patients over 60 kg.

.

*Results*
Perioperative administration of the Non-Steroidal Anti-Inflammatory Drug (NSAID) ketorolac, a common surgical anti-inflammatory analgesic, was associated with significantly superior disease-free survival in the first few years after surgery. The expected prominent early relapse risk peak is all but absent in the 2010 ketorolac data (fig. 7). The few events in the ketorolac group show a small bump in the first 10 months and then slowly rising until the $4^{th}$ year when follow-up of this series ends. After 24 months the ketorolac group hazard rate pattern is indistinguishable from the corresponding pattern for the no-ketorolac group. The updated analysis presented in fig. 8 shows that the benefit appears in the 9-18 month hazards and is of magnitude 4 – 6 fold, consistent with the early report. Specifically in that 9 month period there are 3 relapses in the ketorolac group compared to 15 in the no-ketorolac patients.

Even with the insight of simulations, it is sometimes impossible to determine with certainty what happened to each of the various relapse modes in a particular report. However in this case it appears that perisurgical ketorolac is associated with a dramatic reduction of the recurrences that, according to the proposed model, are related to surgery-induced metastatic activity. If this observation holds up to further scrutiny, it could mean that the simple use of this safe and effective anti-inflammatory agent at the time of surgery might eliminate most early relapses.

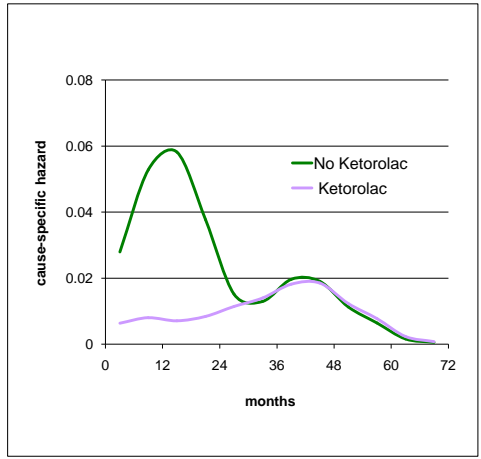

**Fig. 7** Forget et al [40] data from *Universite catholique de Louvain* in Brussels, Belgium. Relapse hazard is shown for mastectomy patients given ketorolac or not. Data are smoothed as indicated for fig. 1.

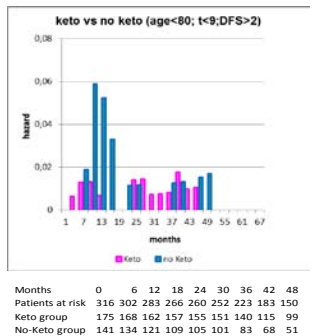

**Fig. 8** Forget et al data were updated September 2011 and shown in hazard form but not smoothed as in fig. 7. Patient data are presented in the table. Patients included in this figure were less than 80 years of age, tumor less than 9 cm diameter and disease free survival greater than 2 months. It can be seen that relapses in months 9 -18 accounted for the major difference between ketorolac and non-ketorolac patients.

*Discussion*

We knew that some intervention starting before surgery would be needed to prevent surgery-induced tumor activity but what could explain the Forget et al data? Published along with the original Forget et al study, an outline of a number of possible effects of surgery and anesthesia on cancer growth was presented by Gottschalk et al [47]. These include stress, immunosuppression, pain, transfusion, inflammation, hypothermia, and a few others. In view of the extensive literature discussing connections and correlations between cancer growth and inflammation, our interest was drawn toward inflammation as possibly a key metastasis producing process.

Inflammation is the body's response to tissue insult. When tissue is damaged, either by physical trauma or by pathogen, a complex cascade of events is triggered [48]. Numerous inflammatory cells and complexes collaborate to attack the invading pathogen, clear debris, reconstitute the extracellular matrix and assist in the proliferation and transfer of healthy cells to the target site. The inflammatory response is the essential and inevitable part of the repair process and a natural defense reaction to trauma. The severity, timing, and local character of any particular inflammatory response depends on the cause, location and site of the area affected, and host's condition [49,50]. The inflammatory response can be intensified by mast cells which release histamine, which then markedly increases the permeability of adjacent capillaries.

Balkwill et al write that if genetic damage is the "match that lights the fire" of cancer, then inflammation is the "fuel that feeds the flames" and that inflammation affects both the survival and proliferation of already initiated cancer cells [51]. Inflammation is a significant component of the tumor microenvironment.

Inflammatory oncotaxis, a term used to describe tumor growth at a site of inflammation, has long been occasionally seen in persons with known or occult cancer and who have local trauma [52]. A clinical observation published in 1914 when it was more common for persons to walk around with known cancer stated: "The localization of secondary tumors at points of injury has been so often remarked upon that it is unnecessary to cite specific instances. The cause for the phenomenon is unknown." [53]. As another example, El Saghir et al published a case report for a smoker with unresectable non-small cell lung cancer who had minor head trauma and a 7 cm diameter tumor grew there in 30 days [54]. We have studied and commented on this case in some detail [55].

Martins-Green et al studied an avian system in which a virus is the carcinogenic agent [56]. When newly hatched chicks are given injections of Rous sarcoma virus, a tumor develops only at the site of injection unless a wound is made a distance away from the primary tumor where a tumor develops at the site of wounding. They found that when inflammation was inhibited, tumors were also inhibited; when inflammation could not be stopped, tumors developed as before.

Transient inflammation after surgery to remove a primary cancer can be both local and systemic [57]. In a colon cancer study, Pascual et al measured the proinflammatory cytokine interleukin-6 (IL-6) in serum prior to surgery and in peritoneal fluid during surgery to establish baseline IL-6, and again at 4, 12, 24 and 48 hours and at 4 days after surgery to determine a temporal trend. They found levels of IL-6 in serum at approximately 1/300 of the concentrations seen in peritoneal fluid. Extrapolating their data, it would seem that levels in serum would gradually return to baseline in a week or so.

There are data associating primary surgery and transient inflammation for breast cancer from Chow et al [58] and from Perez-Rivas et al [59].

Chow et al were studying the effect of clarithromycin on acute systemic inflammation after mastectomy in 54 patients. They measured IL-6, C - reactive protein (CRP), and tumor necrosis factor-alpha (TNF-alpha) in peripheral blood daily from the day before surgery to 3 days afterwards. These markers comprise an inflammation panel that is found useful in renal disease [60]. In both intervention and control groups Chow et al found no particular change in TNF-alpha but 50 – 60% increase in both CRP and IL-6 for 2 or 3 days after surgery and seem to be heading back to normal by the last measurement. Also of note, leukocytes increased 25-30% and platelets decreased 10% with the same temporal pattern as seen for IL-6 and CRP.

Perez-Rivas et al focused on the differential impact of breast surgery on the serum profiles of early breast cancer patients and healthy women. Samples were collected prior to and 24 hours following breast surgery. They found that surgery increased the concentration of several proteins (Colony Stimulating Factor (CSF1), THSB2, IL-6, IL-7, IL-16, Human Epidermal Growth Factor Receptor 2 (HER2), Fas Ligand (FasL) and Vascular Endothelial Growth Factor B (VEGF-B)) in the overall population that include angiogenesis promoters and markers of inflammation. They report high velocity of change in IL-16 and VEGF-A after surgery for invasive disease. They suggest IL-16 is involved in escape from dormancy.

**There are a number of possible mechanisms for a post-surgical inflammatory reaction effect on tumor growth.**

Figure 9 shows a schematic description of what we suspect to be some of the mechanisms how transient systemic inflammation governs metastatic relapse from early breast cancer.

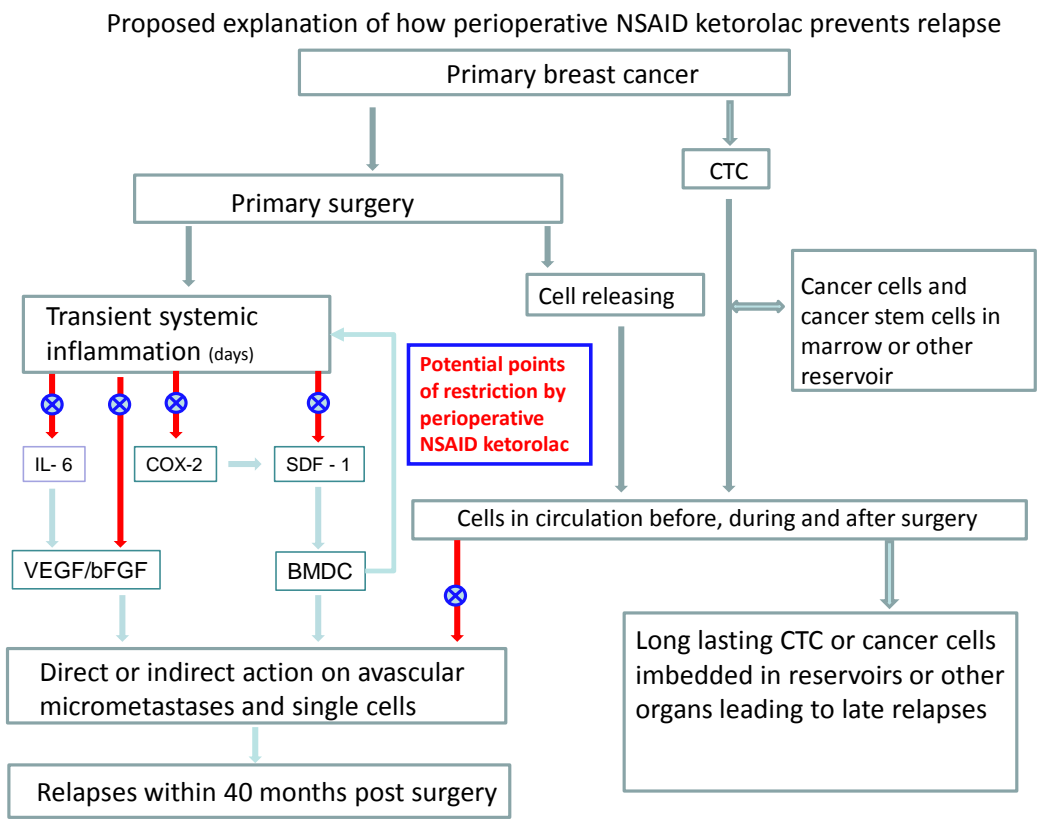

Fig. 9 Symbolic description of proposed explanations for Forget et al data. Early relapses are assumed to be related, at least in part, to the inflammatory process due to primary tumor surgical removal, directly or indirectly eliciting peritumoral endothelial cell and single cell proliferation. A few possible mechanisms are explained. A)

Angiogenic factors, like VEGF and bFGF, are directly released by degranulated platelets or even produced via IL-6; B) Bone marrow derived CXCR-4 positive cells, acting both on tumor foci and on the inflammatory process, are mobilized by SDF-1 directly released or even produced via COX-2. Perioperative ketorolac would restrict both endocrine and cellular pathways, thus impairing the metastatic process. CTC refers to circulating tumor cells.

There are a number of reports that inflammation can spur angiogenesis and tumor growth [49,51,61-63]. This could activate growth of dormant single cells or avascular micrometastases resulting in early relapses. In particular, decreasing the inflammatory response to the surgical maneuver could restrict the angiogenesis switch. A few hypothetical mechanisms can be put forward, including the following.

At steady state conditions in adult mammals, most endothelial cells are quiescent and are believed to contribute to organ homeostasis and tumor dormancy [64]. However, in response to inflammation the upregulation and release of factors stimulating endothelial cells to proliferate could also induce endothelial cells to secrete specific cytokines that reciprocally support the regeneration of normal and malignant stem cells. The metastatic process is believed to be supported by tumor stem cells, which are able to reproduce the cancer progeny. Tumor stem cells, as normal stem cells, require a supporting "niche", i.e., a subset of tissue cells and extracellular substrates defining a specialized microenvironment that is able to modulate the stem cell function (quiescence or proliferation). The occurrence of a metastatic "vascular niche" where endothelial cells play a main role and where an angiogenesis dependent dormancy could result from the cross-talk between tumor cells and endothelial cells (perhaps by regulation of the Notch signaling) has been suggested [65]. If cancer stem cells need to interact with a vascular niche to express their potential, it is reasonable that the latter, under an angiogenic spike by the surgical approach to primary tumor, may appreciably contribute to dormancy interruption [66]. If so, reducing inflammation could result in impairment of the dormant foci wake up process.

Additionally, circulating tumor cells could modify their phenotype secondary to an inflammatory stimulus. When not only analysed for their number, but also for their expression profile linked to activation and ability to adhere, a prerequisite for metastasis formation, circulating tumor cells can modify the expression of nanog, a sign of stem cell properties which enables the cells to self renew and grow. Other factors, like EpCAM, her2/neu, and the adhesion molecule vimentin are other well-known risk factors of proliferation possibly influenced by inflammation [67,68].

Tissue lesions induce mobilization of bone marrow derived cells that are capable of responding to chemo-attractant signals from various organs, where they undergo a homing process and where they release several chemokines [69]. This phenomenon is prominent during neovascularization of wounded tissues via direct or paracrine activity inducing capillary formation. A common basis of the above-mentioned processes is cell trafficking [69]. Indeed, while the intravascular dissemination of normal stem cells is essentially passive, mobilization from their usual niche and homing in a given tissue is regulated by specific signals. Hematopoietic stem cells, for example, express the chemokine receptor CXCR4 and selectively respond to SDF-1α. The SDF-1/CXCR4 axis is a main regulator of the normal cell trafficking underlying the tissue homeostasis. It is also involved in tumor cell trafficking as CXCR4 overexpression is known in more than 20 human tumor types, including ovarian, prostate, esophageal, melanoma, neuroblastoma, and renal cell carcinoma [70dom]. It is, therefore, reasonable to hypothesize that NSAIDs may interfere with SDF1 levels via the pathway COX-2 → PGE → SDF-1, thus resulting in impairment of processes underlying metastasis development.

Even if an NSAID class effect is plausible, a specific effect of ketorolac remains possible. As already stated, whereas all NSAIDs act against the growth of tumors, they are probably not equivalent for this antitumoral effect [40]. Alternative targets, such as the tumor-associated NADH oxydase (tNOX), are possibly involved in this anticancer effect. The existence of tNOX explains the fact that some cancer cell lines lacking COX-2 respond to certain NSAIDs but not to others, suggestive of additional COX-2 independent antitumor activities [71].

Another possible explanation for the lack of surgery-induced angiogenesis when ketorolac is used involves inflammation induced platelet degranulation and that platelets are known to sequester angiogenesis regulating proteins including VEGF [72]. This is especially interesting in view of Chow et al findings that platelets decrease by about 10% in the few days post-surgery. There is also a report that NSAIDs are antiangiogenic and another report that transcript of stem cell marker CD133 that is correlated with poor prognosis in a number of solid tumors was lower in patients treated preoperatively with NSAIDs [73,74].

It is well established that many cancer patients have circulating tumor cells [75-77] and there are cells released as a result of surgery [78]. Camara et al data show a surge in circulating epithelial cells after primary breast cancer surgery, but intriguingly, that surge occurs 3-7 days after surgery. Such a delayed increase in what may be circulating tumor cells after breast cancer surgery was also reported by Daskalakis et al [79]. This phenomenon recalls the surge of CD34+ progenitor cells 3-5 days after tissue damage (e.g., myocardial infarction) [80]. Also it has been recently reported in an animal model, where mice with subcutaneous implantation of Lewis lung carcinoma were subjected to an operative injury, that surgery induced the release of cytokines/chemokines and mobilized bone marrow–derived cells (BMDC) [81]. These mobilized cells were then recruited into tumor tissue with concomitant enhancement of angiogenesis, thereby accelerating tumor growth. Furthermore, blocking recruitment of bone marrow stem cells by disrupting SDF/CXCR signals completely negated the accelerated tumor growth. Many questions arise. Are these surged cells reported by Camara [78] shed or spilled into circulation during surgery and if so why are they delayed by a few days? Or perhaps are these cells released from the bone marrow as part of the programmed wound healing process? Is this a connection to Dvorak's comment that cancer is" wound healing gone awry" [82]? What exactly controls this effect? Can a perioperative NSAID stop this process?

It further suggests that tumors may share physiological mechanisms with normal tissues and, moreover, that inhibiting the inflammatory process might reduce late metastases as well as can be seen in a recent report of daily use of aspirin [83]. Interestingly, the benefit of daily aspirin does not appear until after 2 years of use. This would be consistent with the possibility that late relapses are the result of late inflammation driven events that induce single cell growth and that result in relapses approximately 30 months hence. If late relapses were the result of surgery-induced angiogenesis, we would expect to see a benefit of daily aspirin at 10 or so months after starting.

Blood flow in capillaries is only 0.05cm/sec [84] which would make leaky capillary venules a very efficient way for circulating tumor cells to enter tissue. It may be that what we previously called dormant single cells induced into metastatic growth were at least in some cases residing not at the site of eventual relapse. Rather, circulating tumor cells or BMDC released into circulation by a host response to surgery in an inflammatory environment extravasate, resulting months later in a metastatic tumor. Circulating tumor cells are a reality. Surgical induction of inflammation is universal. Capillary leakage is enhanced by inflammation. It is thereby logical to expect that an effective perisurgical anti-inflammatory strategy may affect surgery-induced and possibly angiogenesis-mediated cancer spread.

The metastatic process is highly inefficient. A clonal malignant cell injected into the circulation has approximately 0.0001 probability to result in a growing metastatic site [85]. Inflammation bypasses the need for extravasation through an intact vessel wall and also provides growth factors to the microenvironment. We estimate the metastatic seeding process is amplified 100 fold during the few days or weeks after primary surgery.

Are the missing early relapses never to happen or are they merely postponed to become late relapses? Whatever their source and shedding timing, cancer cells in circulation may have half-life of a few days or less. Unless injected into more hospitable surroundings such as tissue, these cells will likely harmlessly die off. These data and our analysis suggest that at least for some patients the early relapses apparently avoided in the Forget et al data do not show up later.

Lastly, the reduced recurrence risk for patients receiving perioperative NSAID may be attributed, at least in part, to the reduced usage of opioids for pain management with ketorolac [86,87]. It cannot be excluded that all the above mentioned mechanisms could act together resulting in relapses within the subsequent few years.

**TNBC and early relapses – possibly an ideal group for testing perioperative ketorolac.**
We now turn our attention to methods of testing this new hypothesis. Animal studies would be very important, however in view of the analyses and data presented we think this should be tested prospectively in a clinical trial. The next question that arises is what patient group would be a good candidate for a trial. Most breast cancer clinical trials, at least in the US, focus on distinct patient subgroups based on recurrence risk levels. The triple negative subgroup attracted our attention for several reasons [29]. Lacking markers for HER2, Estrogen or Progesterone receptors that strongly suggest that there is benefit of targeted therapy, triple negative breast cancer (TNBC) is looked upon by clinicians as a "bad tumor" with high recurrence rate in spite of adjuvant chemotherapy. That pessimistic viewpoint seems justified since TNBC has 12% incidence but accounts for approximately 20% of mortality in breast cancer.

We had access to a triple negative breast cancer data base from Milan that we analyzed with our hazard methods. The relapse hazard (fig. 10) looks remarkably similar to the no-ketorolac group in the Forget et al study shown in fig. 7. Triple negative breast cancer therefore appears to be the ideal study group with which to test benefit of perioperative ketorolac in a clinical trial.

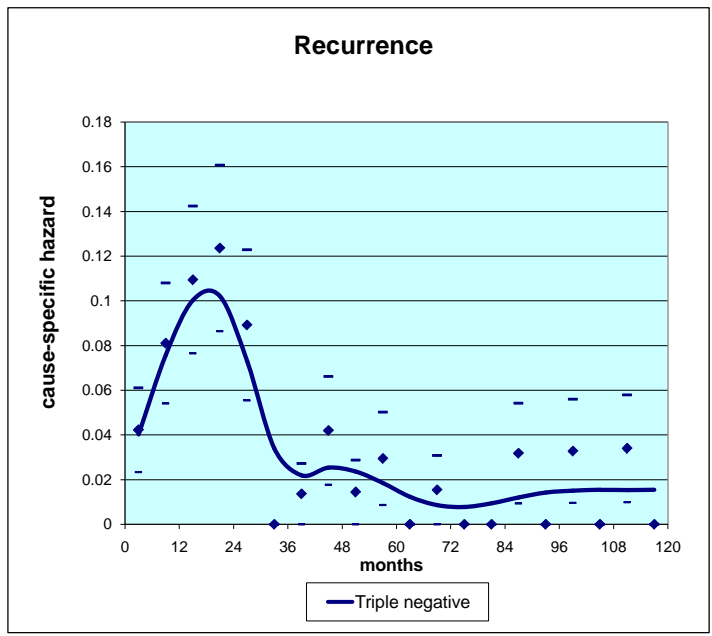

**Fig. 10** Hazard data from a Milan database for 121 TNBC patients with 10-11 years follow up. There are 50 relapse events within 5 years of surgery. The general similarity of these hazard data to Forget et al data for no-ketorolac patients seen in Fig. 7 leads to the suggestion that TNBC may be the ideal study group with which to test perioperative ketorolac.

The incidence of TNBC is 12% in US population (as mentioned), 25% among African Americans, and 25% to 35% among patients from India and Korea [88-90]. (There may be other as yet unexamined groups also with high incidence of TNBC.) Locations with relatively high incidence of TNBC would be ideal places to conduct a clinical trial in order to make it easier to show an improvement in early relapse.

As noted by Wallace et al, the racial disparity in breast cancer outcome is due primarily to deaths within the first few years after diagnosis providing an additional motivation to test at the earliest opportunity what we report here [91]. That would be consistent with the information just noted.

**Concerns about bleeding complications.**

One of the issues related to the perioperative use of ketorolac has been concern about bleeding complications. What is the evidence, if any, regarding the occurrence of increased blood loss and its clinical significance after a single or limited number of doses of ketorolac when administered during the perioperative period?

This topic has been recently addressed in an editorial by White, Raeder and Kehlet, accompanying a meta-analysis of De Oliveira et al [92,93]. In the meta-analysis, the authors noted that the combined effect did show a statistically significant increase in bleeding with ketorolac compared with placebo. This effect was however shown only in two studies focusing on surgeries with "raw" surface areas (adenotonsillectomy and major orthopedic surgery) and without any additional red blood cell transfusion needed, questioning the clinical significance in other surgeries.

In breast surgery, a recent retrospective study in major plastic breast reconstructive surgery (mammoplasty) reported a greater likelihood of requirement for surgical hematoma evacuation [94]. But, as in surgeries with "raw" surface areas, such major plastic surgeries are associated with greater difficulties in hemostasis than lumpectomy (often performed on a day-case basis) or mastectomy. Two studies in breast cancer surgery prospectively compared ketorolac with placebo. The first did not show any difference in drain output, but is difficult to interpret because ketorolac was administered near the end of surgery (in place of preincisional) [95]. The second showed a statistically significant difference but no clinical implications including no need of transfusion in any group [96].

As a consequence, if the use of ketorolac has been associated with a greater amount of blood loss in a limited number of studies and not in others, the clinical significance remains unknown in breast cancer surgery. If any, it seems to be low since ketorolac has never been associated with greater transfusion need of red blood cells. Studies even tend to report a better functional outcome in the postoperative period with ketorolac, suggesting that the clinical significance of this blood loss could be largely counterbalanced by the advantage of the drug [92,93]. As a consequence, the American Society of Anesthesiologists recommended in their latest guidelines that unless contraindicated, all the patients should receive balanced analgesia, including NSAIDs [97].

**Summary and Conclusions**

Careful analysis of breast cancer recurrences suggests a paradigm where early recurrences, i.e. the majority of adverse events resulting in poor prognosis, are induced by angiogenic switching of avascular micrometastases and single cell activation. Both events are triggered by primary tumor surgical removal.

Results reported by Forget et al analysis of retrospective data, suggesting perioperative NSAID ketorolac significantly reduces early relapses, may be deciphered in the light of this model. Indeed, post-surgical transient systemic inflammation might be the precipitating factor and common denominator for early relapses. In particular, inflammation would be important for angiogenesis induction of avascular distant micrometastases.

Several molecular processes could be involved, either as single mechanisms or concurrently. For example, inflammation induced upregulation and release of factors stimulating endothelial cells to proliferate could also induce endothelial cells to secrete specific cytokines that reciprocally support the regeneration malignant stem cells within the metastatic niche. Or else, as the SDF-1/CXCR4 axis is a main regulator of normal and tumoral cell trafficking, it is reasonable to hypothesize that NSAIDs may interfere with SDF1 levels via the pathway COX-2 → PGE → SDF-1, thus resulting in impairment of processes underlying metastasis development. Another possible explanation involves inflammation induced platelet degranulation, with release of angiogenesis regulating factors including VEGF, which would countered by ketorolac. Lastly, tumor cells released as a result of surgery in the presence of transient systemic inflammation and capillary permeability could also account for succeeding metastatic development.

A few points need further investigations. First, the Forget et al findings need to be confirmed in randomized clinical trials. Such investigations are imperative not only from the scientific point of view but more so for their possible clinical consequences, resulting from the fact that breast cancer mortality could be reduced by 25 to 50% at low cost and toxicity. A subset of patients for a randomized clinical trial should be characterized by unfavorable prognostic factors resulting in early recurrences covering the first 2 to 3 years. We suggest that the best breast cancer population for such a trial may be triple negative breast cancer.

In spite of the fact that breast cancer is known as a disease that runs its course in a decade or more, most of the relevant events resulting in recurrences apparently occur shortly after primary surgery. Investigations focused on events occurring during the first few days and weeks following primary tumor removal are strongly warranted.

Winquist and Boucher describe the lack of innovative new paths moving forward in cancer therapeutics as bleak with improvements often measured in months [12]. The new path outlined here could be a revolutionary break ("Something for nothing" rarely if ever happened in cancer therapy) from the past and should also be explored in other neoplasias. High priority should be given to test this hypothesis as it is implementable regardless of state of socio-economic development because of its low cost. We end by acknowledging the seminal work of friends and mentors Bernard Fisher and the late Judah Folkman.

Acknowledgements: We acknowledge the support of Komen for the Cure Grant: 100484

Abbreviations – TNBC: Triple Negative Breast Cancer; CMF: Cyclophosphamide, Methotrexate, 5-Fluorouracil; CEBH, Commission d'Ethique Biomédicale Hospitalo-Facultaire de l'Université catholique de Louvain; NSAID: Non-steroidal anti-inflammation drug; IL-6: Interleukin-6; CTC: Circulating tumor cells, BMDC: bone marrow derived cells

Conflicts of Interest: M. Retsky has a patent pending for treatment of early stage cancer and on Board of Directors of Colon Cancer Alliance (www.ccalliance.org). No other conflicts of interest reported.

Disclosure: Part of information included in this article has been previously published in *Breast Cancer Research and Treatment.* **2012**, 134(2), 881-888,doi:10.1007/s10549-012-2094-5.